\documentclass[a4paper,11pt,pdftex]{article}

\usepackage[utf8]{inputenc}

\usepackage[dvips]{graphicx}
\DeclareGraphicsExtensions{.pdf, .jpg, .png} 
\graphicspath{{../fig/}}

\usepackage[section]{placeins} 

\usepackage{pslatex}

\usepackage{geometry}

\usepackage{array}
\newcolumntype{A}{>{\centering\let\newline\\\arraybackslash\hspace{0pt}}m{0.22\textwidth}}
\newcolumntype{B}{>{\centering\let\newline\\\arraybackslash\hspace{0pt}}m{0.15\textwidth}}

\usepackage[normalem]{ulem}
\usepackage{ifthen}

\usepackage{bigints}
\usepackage{scrextend}

\usepackage{xcolor}
\usepackage{soul}
\usepackage{epsfig}
\usepackage[numbers,sort&compress]{natbib} 

\usepackage[font=footnotesize,hang]{caption} 
\usepackage{euscript}
\usepackage[bottom]{footmisc}
\usepackage{gensymb} 
\usepackage{siunitx} 

\usepackage{xspace} 
\usepackage[USenglish]{babel} 
\usepackage{microtype}

\usepackage{lineno}

\usepackage[T1]{fontenc}

\usepackage{verbatim}
\usepackage{numprint} 
\usepackage{eso-pic}
\usepackage{subfigure}
\usepackage{multirow}
\usepackage{epigraph}
\usepackage{hyperref} 
\usepackage{cleveref,nameref}

\usepackage{authblk}

\usepackage[useregional]{datetime2}

\usepackage{fancyhdr}
\usepackage{lastpage}
\usepackage{setspace}

\DeclareSymbolFont{Letters}{U}{zeur}{m}{n} 
\DeclareMathSymbol{\psi}{\mathalpha}{Letters}{"20} 

\usepackage{amsfonts}
\usepackage{amssymb}
\usepackage{amsmath}
\usepackage{mathrsfs}

\let\phi\varphi
\let\theta\vartheta

\renewcommand{\sin}{\operatorname{sin}}

\renewcommand{\imath}{\operatorname{i}}

\title{Analog FM free-space optical communication based on a mid-infrared quantum cascade laser frequency comb } 
\date{}

\author[1,*]{Nicola Corrias}
\author[1]{Tecla Gabbrielli}
\author[1]{Paolo De Natale}
\author[1]{Luigi Consolino}
\author[1,*]{Francesco Cappelli}

\affil[1]{CNR-INO -- Istituto Nazionale di Ottica, Largo Enrico Fermi 6, 50125 Firenze FI, Italy  \newline \& LENS -- European Laboratory for Non-Linear Spectroscopy, Via Nello Carrara 1, 50019 Sesto Fiorentino FI, Italy}

\affil[*]{Corresponding authors: \texttt{nicola.corrias@ino.cnr.it}, \texttt{francesco.cappelli@ino.cnr.it}}

\begin{document}

\maketitle
\thispagestyle{fancy}


\begin{abstract}

Quantum cascade laser frequency combs are nowadays well-appreciated sources for infrared spectroscopy. Here their applicability for free-space optical communication is demonstrated. The spontaneously-generated intermodal beat note of the frequency comb is used as carrier for transferring the analog signal via frequency modulation.  
Exploiting the atmospheric transparency window at \SI{4}{\micro \meter}, an optical communication with a signal-to-noise ratio up to \SI{65}{\dB} is realized, with a modulation bandwidth of \SI{300}{\kilo \hertz}. The system tolerates a maximum optical attenuation exceeding \SI{35}{\dB}. 
The possibility of parallel transmission of an independent digital signal via amplitude modulation at \SI{5}{\mega S / \second} is also demonstrated. 

\end{abstract}

\section{Introduction}

The mid-infrared (MIR) region of the electromagnetic spectrum ($\lambda$~\SI{> 3}{\micro \meter}) is of particular interest both from a fundamental and an applied research point of view. Here fall in fact the strongest absorptions related to the ro-vibrational transitions of light molecules of environmental/atmospheric interest such as water, carbon dioxide, methane and nitrogen oxides. From a fundamental viewpoint, the study of the absorption spectra with ultrahigh precision and sensitivity can give insights on fundamental laws of physics~\cite{Modugno:1998,Modugno:2000,Mazzotti:2001,Shelkovnikov:2008,Cournol:2019,Fast:2020,Kortunov:2021}. From an applied perspective, the high-sensitivity detection of such gases through advanced spectroscopy techniques~\cite{Adler:2010b,Foltynowicz:2013} is important for environmental and health monitoring~\cite{Lashof:1990}. The MIR region is not completely covered by atmospheric gases absorptions: Two main transparency windows (one around \SI{4}{\micro \meter} and one around \SI{10}{\micro \meter}) are present. Here the atmospheric transmittance (per km) is > 80\% (with peaks of 99.965\%) and they can be conveniently exploited for free-space optical communication (FSOC)~\cite{HITRANw:2019,HITRANd:2013}. In addition, the attenuation due to the aerosol is also significantly lower compared to visible and near-infrared wavelengths~\cite{Elterman:1964}.  

Since its demonstration in 1994 at Bell Labs~\cite{Faist:1994}, the quantum cascade laser (QCL) has been largely used and appreciated as coherent source~\cite{Faist:2013bo}. QCLs are chip-scale room-temperature current-driven semiconductor-heterostructure lasers where the laser transition takes place between the sublevels of the conduction band. This sublevels structure is given by the semiconductor layers heterostructure that creates quantum wells for the carriers. Since the structure and thickness of the layers are highly engineerable through molecular-beam epitaxy, the levels' structure can be freely arranged. Therefore, in combination with the possibility of using different families of semiconductor materials, the emission wavelength can be engineered in a wide range, spanning from the MIR ($\lambda$~\SI{> 3.5}{\micro \meter}) to the THz~\cite{Beck:2002,Kohler:2002}. Due to the very short laser transition lifetime ($< 1$~ps) QCLs can be modulated at high speed (GHz and above)~\cite{Hinkov:2016,Hinkov:2019}. Their high-power emission ($\sim$~W)~\cite{Hinkov:2013} and current driving make them ideal candidates for infrared spectroscopy~\cite{Consolino:2018a,Borri:2019a,Galli:2013b,Galli:2016b,Santagata:2019}. 
QCLs can operate both in single longitudinal mode \cite{Faist:1997} and in frequency-comb regime \cite{Hugi:2012,Burghoff:2014,Rosch:2014,Faist:2016}. Indeed broadband QCLs \cite{Riedi:2013,Riedi:2015} can emit frequency combs (QCL-combs) due to the high third-order nonlinearity characterizing the active region which enables a four wave mixing (FWM) parametric process within the waveguide~\cite{Friedli:2013}. FWM correlates the generated modes~\cite{Khurgin:2014,Henry:2018,Opacac:2019,Burghoff:2020,Khurgin:2020} establishing a fixed phase relation among them and ensuring a coherent frequency-comb emission~\cite{Cappelli:2016,Cappelli:2015,Burghoff:2015,CappelliConsolino:2019,Consolino:2019a}, which proved to be successfully exploitable for dual-comb spectroscopy~\cite{Villares:2014,Consolino:2020a}. An intermodal beat-note signal (IBN) is associated with the frequency-comb emission, given by the beatings between first-neighbor modes. This signal, falling in the range 5-20~GHz, is determined by the waveguide length and effective refractive index, and it can be as narrow as few hundreds of Hz or less due to the high overall coherence proper of QCLs and QCL-combs~\cite{Vitiello:2012,Cappelli:2015,Shehzad:2020}. IBNs can be easily modulated by modulating the laser's driving current~\cite{Consolino:2019a} or by other means (e.g. optical modulation~\cite{Consolino:2021b} or modulating the current passing through an integrated resistive heater~\cite{Gurel:2016}). 

The single-mode emission of QCLs has already been exploited for FSOC relying on amplitude modulation (AM)~\cite{Capasso:2002,Pang:2020}. Regarding QCL-combs, the possibility of exploiting their IBN for transmitting an analog signal in the microwaves in a wireless-broadcasting fashion has been recently demonstrated~\cite{Piccardo:2019a}. This configuration is mainly useful for indoor short-range communication.

In this work we demonstrate the possibility of exploiting the IBN of a QCL-comb for carrying the signal with frequency-modulation (FM) over the optical channel corresponding to the atmospheric transparency window around 4~$\mu$m and we characterize the performance of the communication system. Moreover, we demonstrate that an independent digital signal can be encoded on the same optical channel with the same device via AM. This configuration is particularly well-suited for long-distance point-to-point transmission in the atmosphere.

\section{Methods and Discussion}

\subsection{Experimental setup}

\begin{figure}[!htbp]
\begin{center} 
\includegraphics[width=1\textwidth]{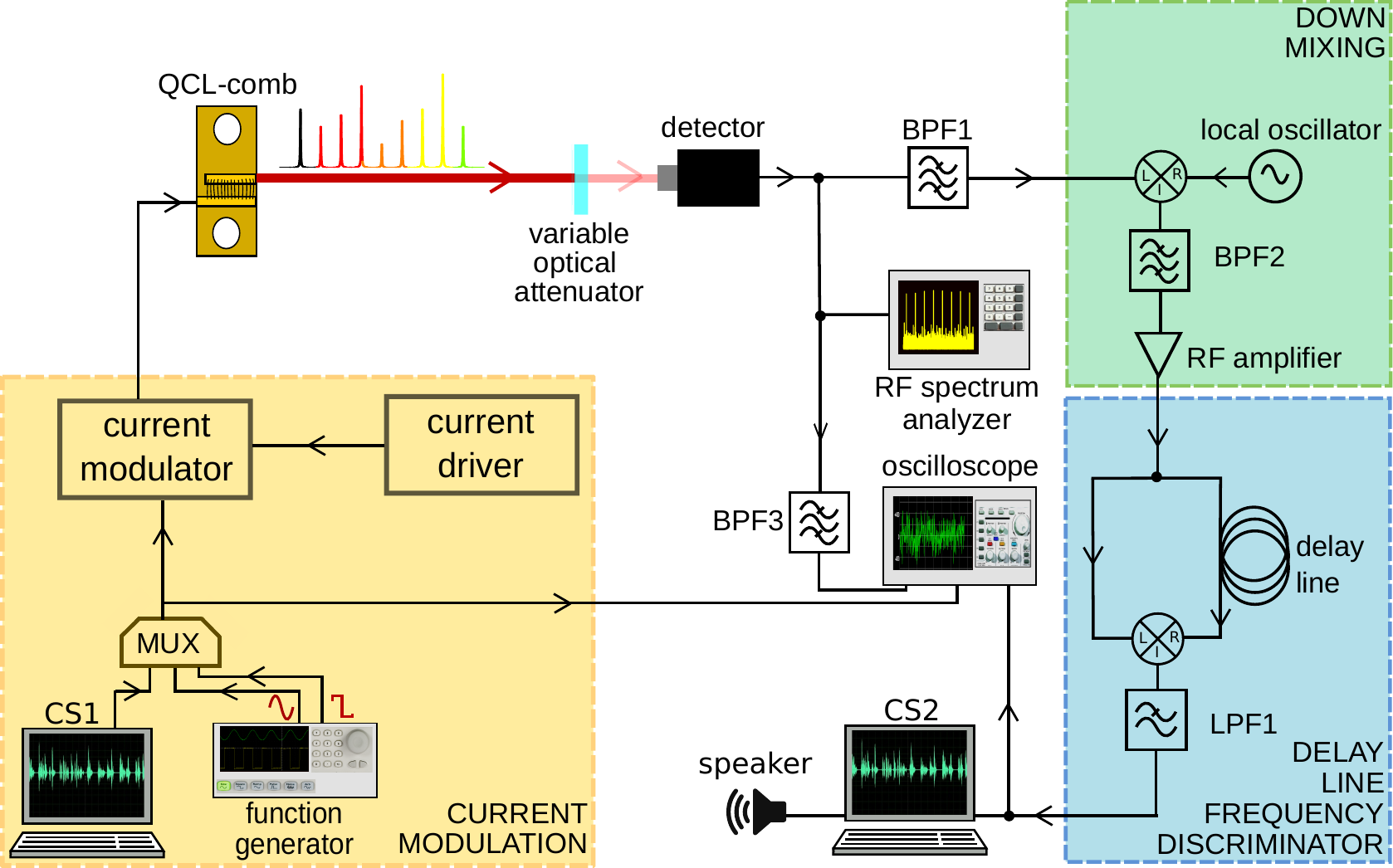} 
\caption[]{Experimental setup of the system for the optical free-space communication. The source is a MIR QCL-comb (top left). The laser's driving system and the signals encoding system is highlighted in the yellow box. On the right, the components devoted to the detection, demodulation and analysis of the signals is represented. BPF: band-pass filter. LPF: low-pass filter. CS: computer system. } 
\label{fig:QCL-comb_comm_setup}
\end{center}
\end{figure}
The experimental setup used for demonstrating and characterizing the optical communication system is sketched in \cref{fig:QCL-comb_comm_setup}. The laser source utilized for the transmission is a QCL-comb designed and fabricated at ETH Z\"urich. The device is a buried-heterostructure Fabry-P\'erot QCL able to operate in continuous wave at room temperature \cite{Cappelli:2016,CappelliConsolino:2019}. Its emission is centered around \SI{4.70}{\micro \meter} and the mode spacing ($f_s$) corresponding to the intermodal beat note (IBN) is about \SI{7.06}{\giga \hertz}. Further details about the laser source (LIV curves and laser spectrum) are given in the \nameref{sec:supplementary}. The working temperature of the laser for this experiment varied between 12 and \SI{20}{\celsius}. The laser radiation propagates in the free space for \SI{2.0}{\meter} and it is sent through a variable attenuator to a fast thermoelectric-cooled photovoltaic MCT detector (PVI-4TE-8-0.5x0.5-T08-wBaF2-35 by Vigo), operating in the spectral range from \SIrange{3}{9}{\micro \meter}. The detector features a two-stages amplification system: the first stage is a high-speed transimpedance, the second stage is an AC-coupled preamplifier. The bandwidth (\SI{-3}{\dB}) is \SI{700}{\mega \hertz}. 

\begin{figure}[!htbp]
\begin{center} 
\includegraphics[width=0.5\textwidth]{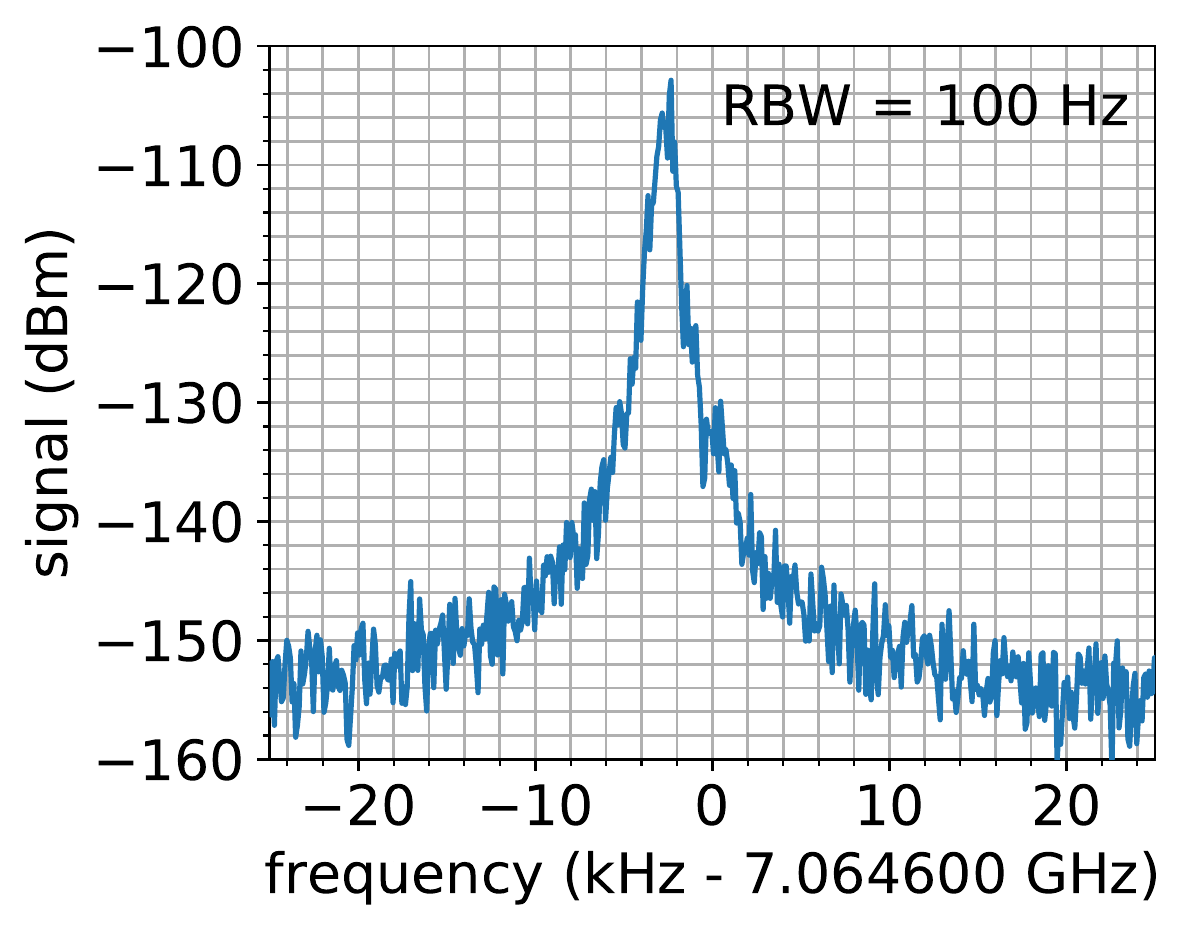}
\caption[]{Spectrum of the QCL-comb IBN in free-running operation at $T = \SI{12}{\celsius}$ and $I = \SI{730}{\milli \ampere}$ detected by the MCT detector and acquired with a spectrum analyzer. } 
\label{fig:IBNzoom}
\end{center}
\end{figure}

The electronic part of the setup consists mainly of two subsystems. The first one, dedicated to signal generation, modulation and transmission, is composed of the signal generation system and the laser current driver. The laser is powered by an ultra-low-noise current driver (QubeCL, by ppqSense) featuring a module for current modulation, that is fed with the signal to be transmitted. For demonstration purposes, the laser current is modulated with sine waves, square waves and analog audio signals generated via a function generator or a computer system (CS1). The signals to be sent to the current modulator are selected via a multiplexer (MUX). The modulation of the laser current induces a frequency modulation of the IBN of the laser. The IBN is out of the nominal bandwidth of the detector but can still be well detected anyway (see \cref{fig:IBNzoom}). 

The second subsystem of the electronic part is devoted to the extraction of the original signal from the detected IBN via a demodulation process. At first, the IBN is down-mixed from \SI{7.06}{\giga \hertz} to \SI{54.6}{\mega \hertz}, then it is band-pass-filtered (BPF2, band 5~MHz) and finally amplified (green box in \cref{fig:QCL-comb_comm_setup}). Afterwards, a delay-line frequency discriminator (DLD) -- optimized for operating at frequencies around the down-mixed carrier -- performs the demodulation of the IBN~\cite{Faith:1988} extracting the original signal (see the \nameref{sec:supplementary} for a description of the DLD). After the DLD, a low-pass filter removes the down-mixed carrier (blue box in \cref{fig:QCL-comb_comm_setup}). Finally, the demodulated signal is sent to an oscilloscope to be compared with the original signal and to a second computer system (CS2) to be further analyzed.

\subsection{Intermodal beat-note characterization}

\begin{figure}[!htbp]
\begin{center} 
\includegraphics[height=0.35\textwidth]{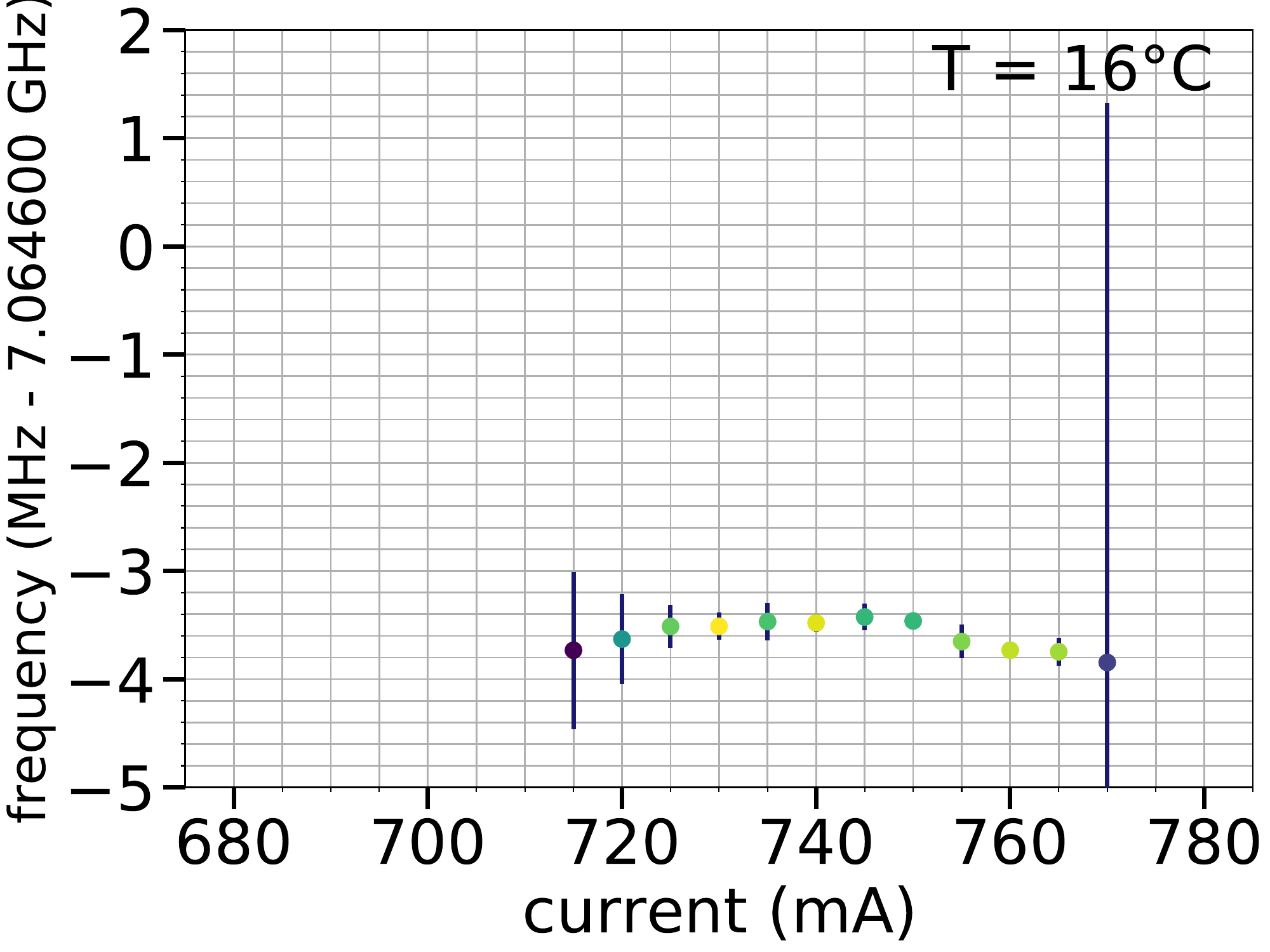} 
\includegraphics[height=0.35\textwidth]{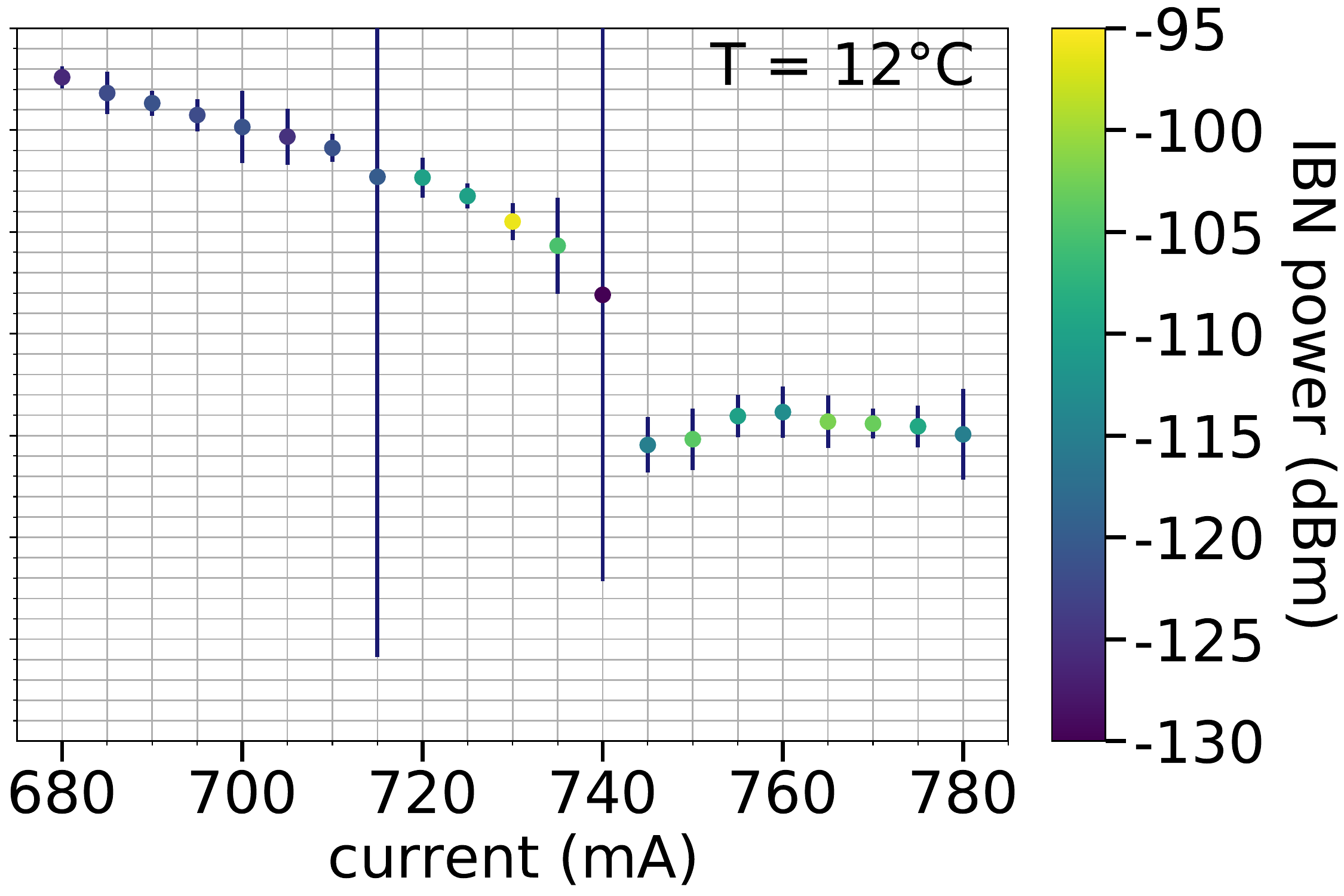}
\caption[]{Frequency tuning with bias current of the IBN for two different temperatures of operation of the QCL. The bars represent the width of the IBN, where the values have been multiplied by a factor 500 in order to have visible bars in the graphs. The peak values of the IBN are color coded. } 
\label{fig:IBNtuning}
\end{center}
\end{figure}

In order to perform the best FM analog transmission it was crucial to determine the working conditions of the laser (temperature and driving current) where the IBN is intense and narrow (with the smallest frequency noise) and the FM is optimized, i.e. the working condition where the frequency tuning of the IBN with bias current is large. The analysis is performed at several temperatures in the range from $T = \SI{20}{\celsius}$ to $T = \SI{12}{\celsius}$. For each temperature, a scan of the driving current, acquiring the IBN, is performed. As an example, the IBN response at two different temperature values is reported in \cref{fig:IBNtuning}. 
At $T = \SI{16}{\celsius}$ the IBN is pretty intense and narrow (the error bars representing the width are small) but the tuning with current of the IBN frequency is very small and it is almost zero around $I = \SI{740}{\milli \ampere}$. On the other hand, at $T = \SI{12}{\celsius}$ the tuning is generally large and there are intervals where the IBN is intense and narrow. Finally, we chose $T = \SI{12}{\celsius}$ and $I = \SI{730}{\milli \ampere}$ as working conditions, since here all the considered parameters are optimized (see \cref{fig:IBNzoom} representing the spectrum of the IBN in free-running operation). We remark that these working conditions are definitely compatible with standard operating conditions, i.e. room temperature and moderate driving current. 

\begin{figure}[!htbp]
\begin{center} 
\includegraphics[width=1.0\textwidth]{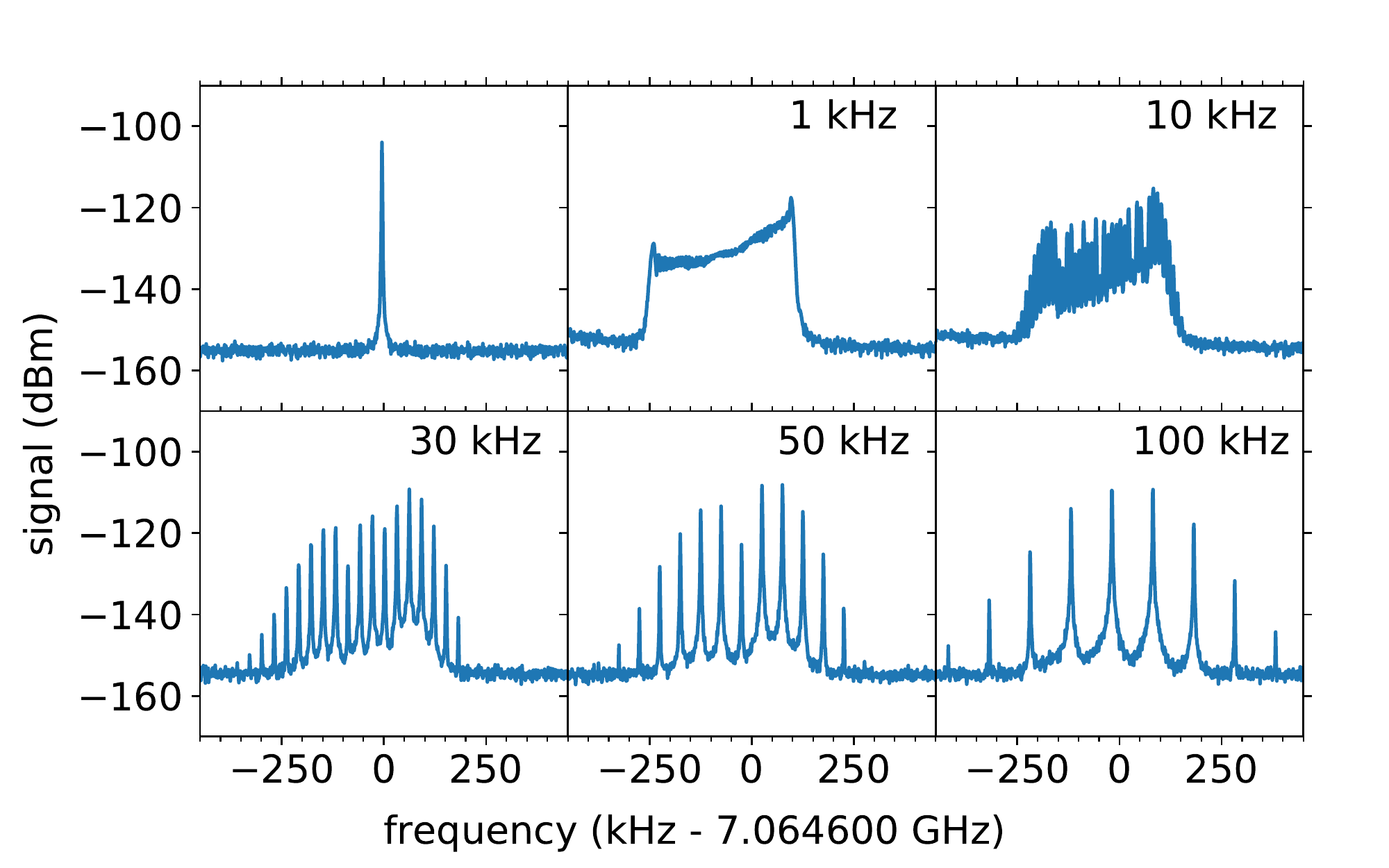}
\caption[]{Spectra of the detected IBN with different sinusoidal frequency modulations acquired with a spectrum analyzer. The QCL-comb operates at $T = \SI{12}{\celsius}$ and $I = \SI{730}{\milli \ampere}$. } 
\label{fig:IBNmodupan}
\end{center}
\end{figure}

Afterwards, a characterization of the modulated IBN has been performed. With the MUX, a sinusoidal signal has been sent to the current modulator and the amplitude of the signal has been chosen as the maximum value tolerated by the laser for preserving the frequency-comb emission. In terms of driving current, the current modulation amplitude is about 1\% of its DC value. The frequency of the signal has been varied to characterize the response of the system. In \cref{fig:IBNmodupan} the IBN modulated at different frequencies is shown. 
From the plots, a peak frequency deviation of about 200~kHz can be inferred. This value is significantly large compared to the linewidth of the IBN (see \cref{fig:IBNzoom}). In these conditions the frequency noise is negligible with respect to the frequency deviation encoding the transmitted signal.

\subsection{Characterization of the received analog FM signal}

\begin{figure}[!htbp]
\begin{center} 
\includegraphics[width=0.48\textwidth]{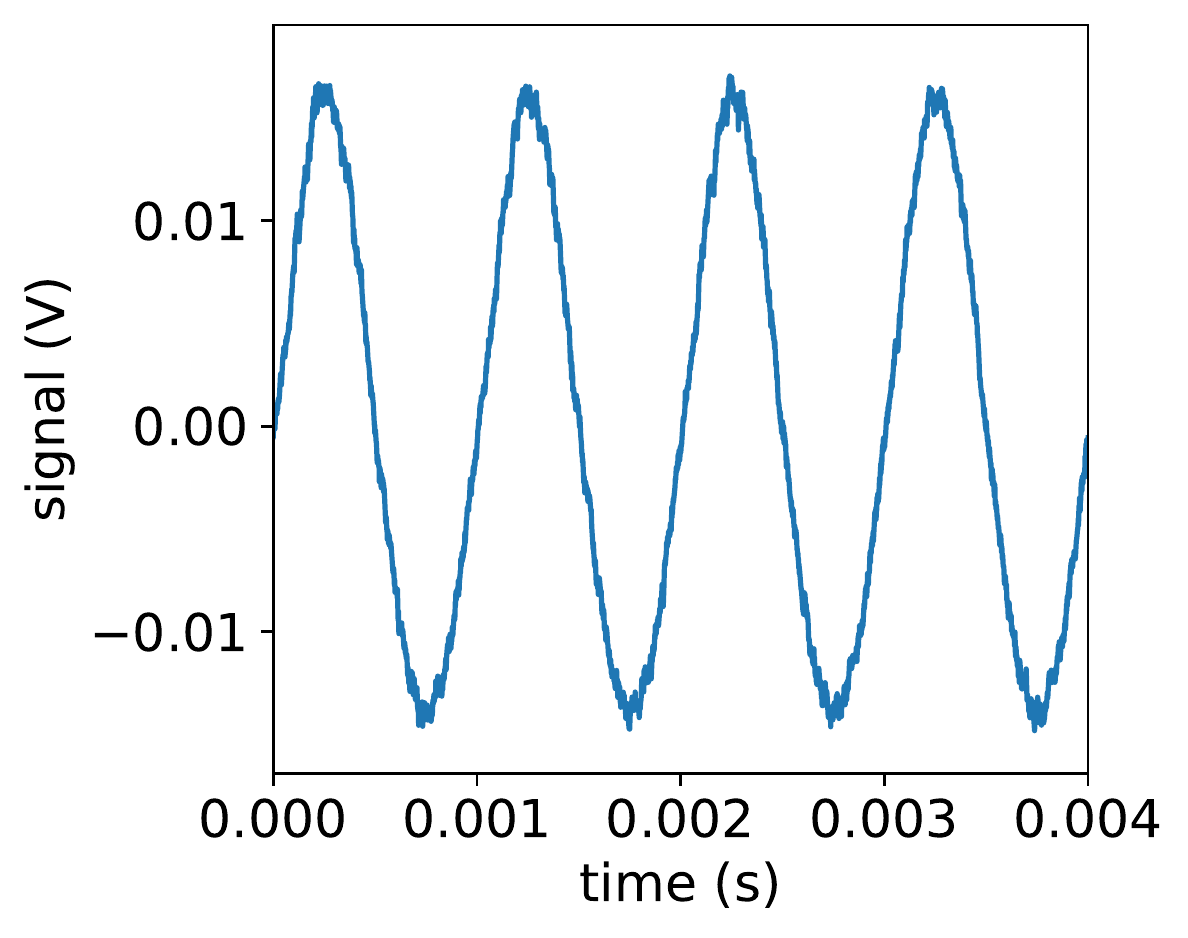}
\quad 
\includegraphics[width=0.48\textwidth]{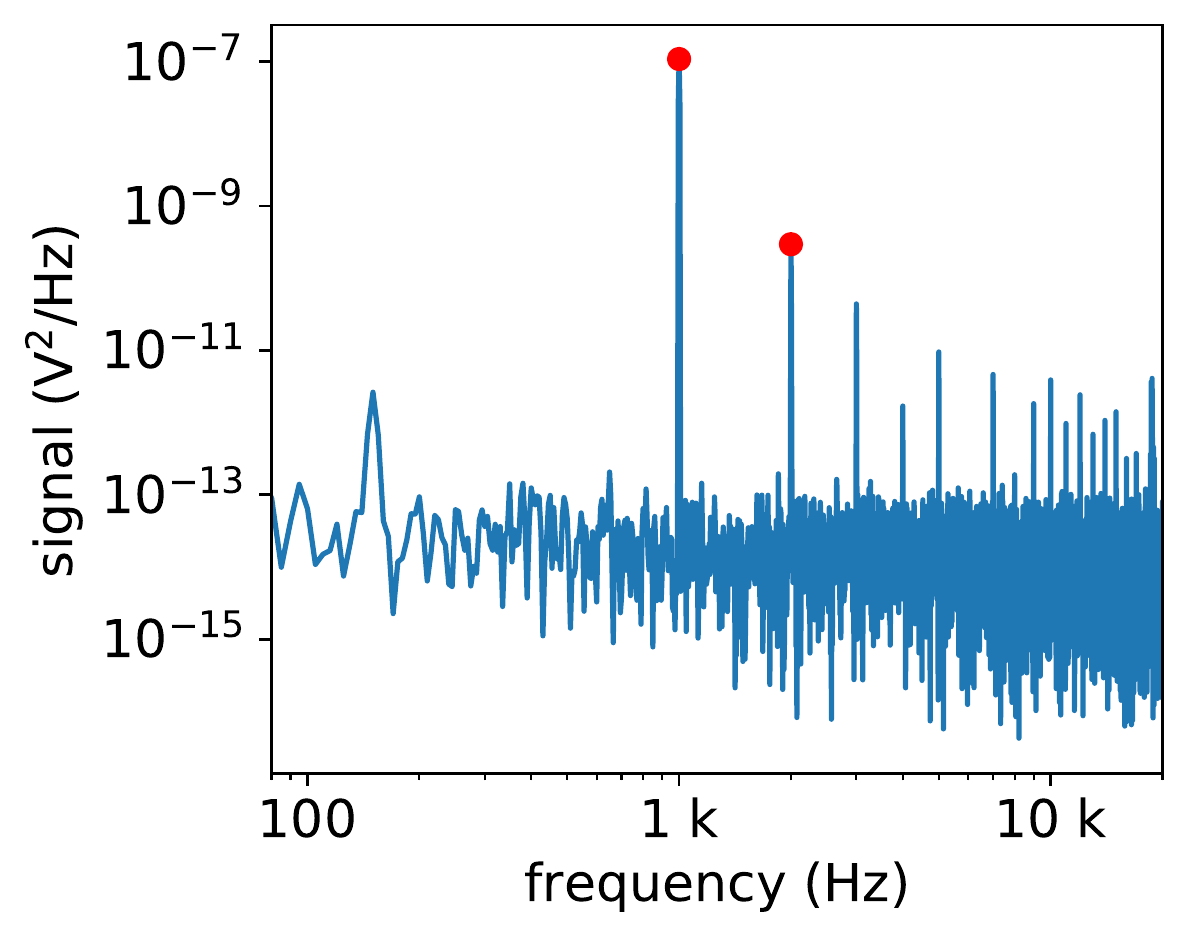}
\caption[]{Left: Demodulated signal at \SI{1}{\kilo \hertz}. Right: FFT of the demodulated signal with the fundamental frequency and second harmonic peaks marked with red dots. } 
\label{fig:rec_sig_FFT}
\end{center}
\end{figure}

\begin{figure}[!htbp]
\begin{center} 
\includegraphics[width=0.60\textwidth]{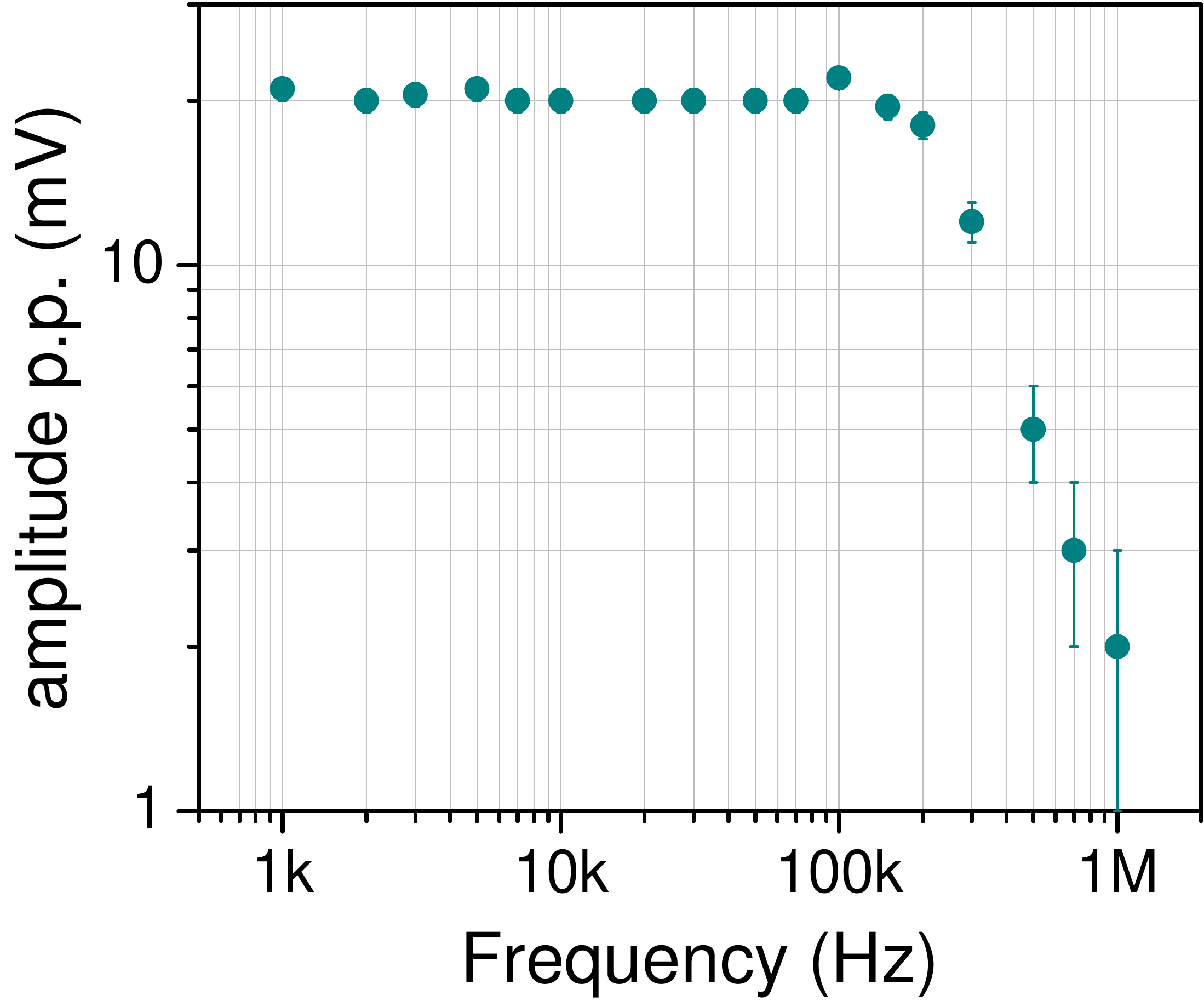}
\caption[]{Modulation response of the system. 
The bandwidth is $\sim \SI{300}{\kilo \hertz}$, perfectly in agreement with the well-known single mode QCLs' frequency modulation bandwidth~\cite{Tombez:2013a}. } 
\label{fig:freqModBand}
\end{center}
\end{figure}

\begin{figure}[!htbp]
\begin{center} 
\includegraphics[width=0.48\textwidth]{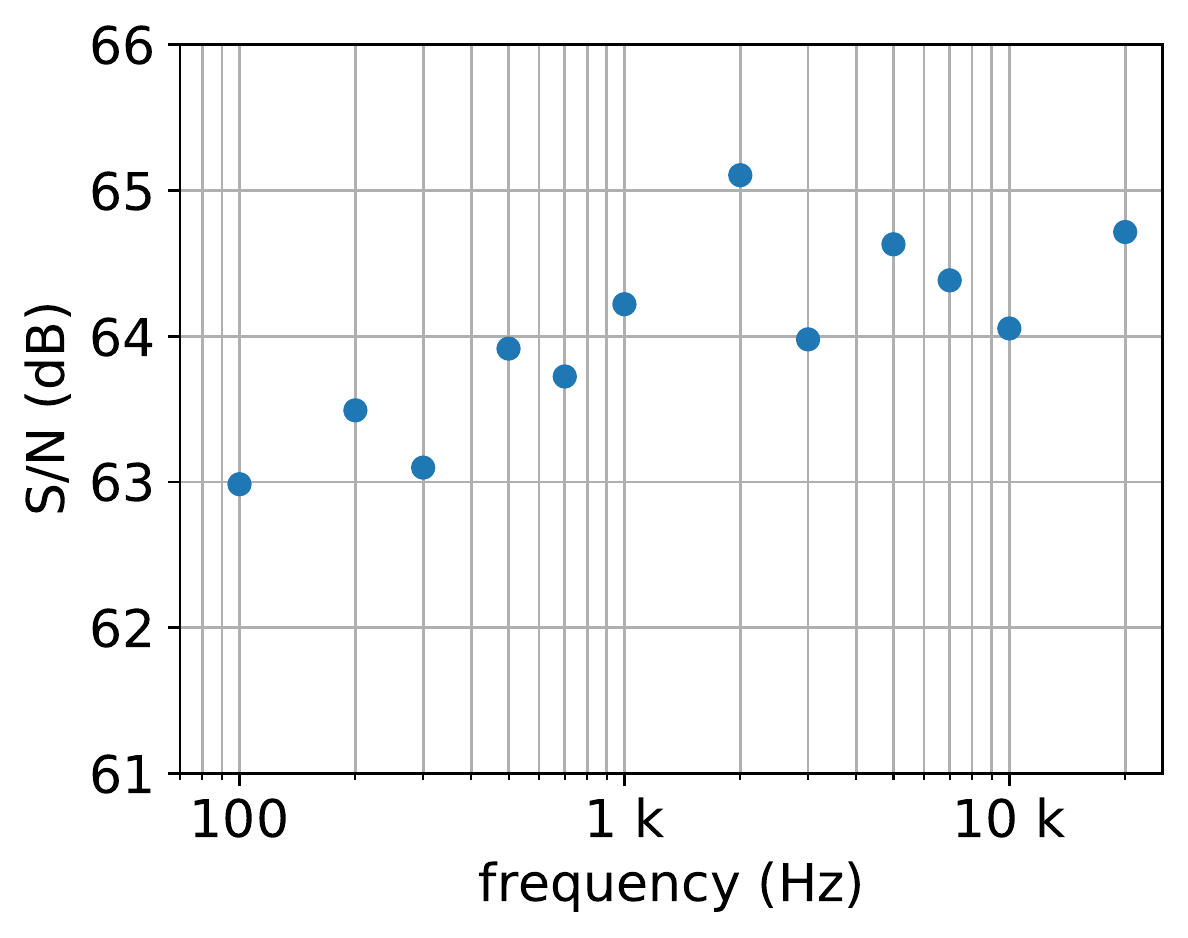}
\quad 
\includegraphics[width=0.48\textwidth]{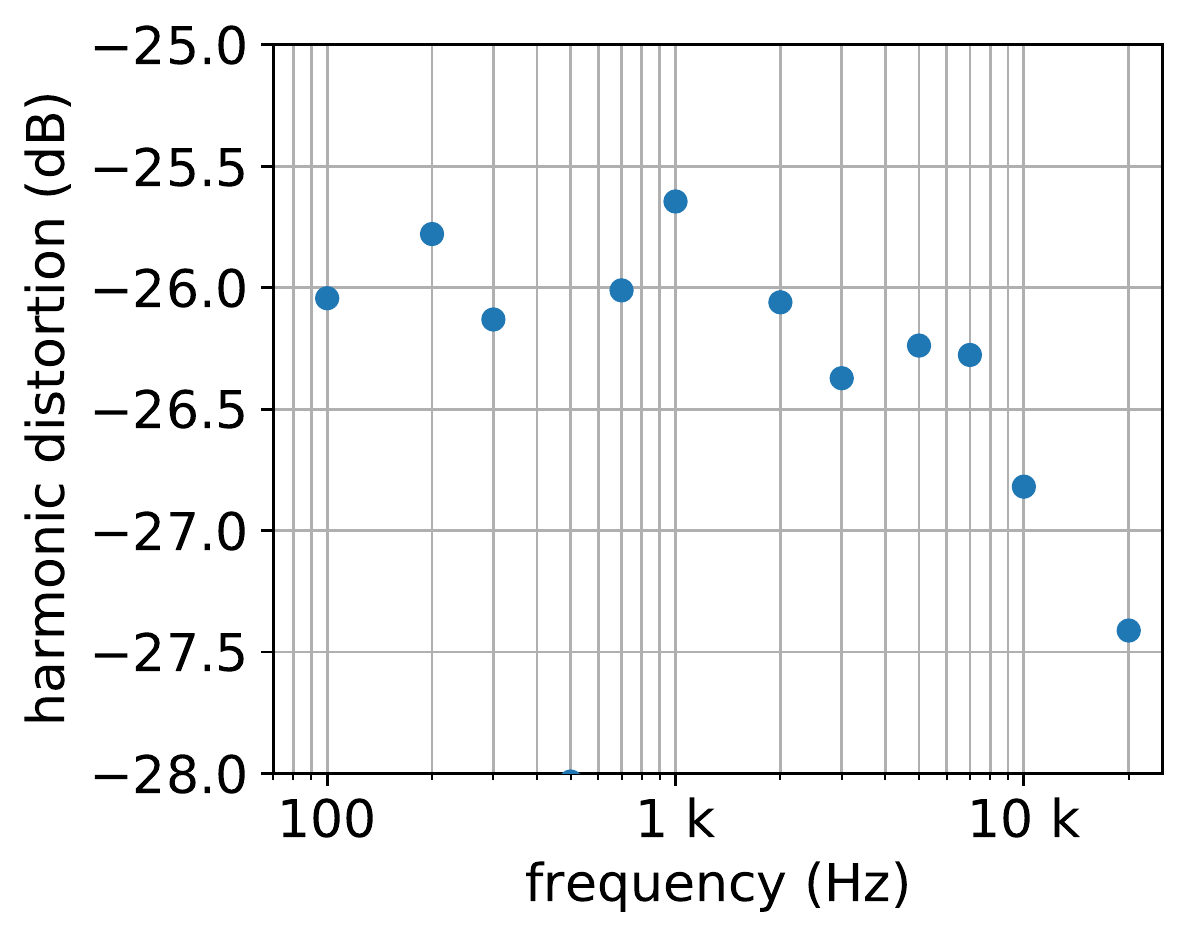}
\caption[]{Left: S/N as a function of modulation frequency. Right: Harmonic distortion as a function of modulation frequency. The harmonic distortion is computed as the ratio between the power of the second harmonic and the power of the fundamental frequency peaks of the received signal. } 
\label{fig:harm_dist_SNRvsfreq}
\end{center}
\end{figure}

For quantitatively estimating the quality of the transmission, the signal demodulated by the DLD has been acquired with an oscilloscope. \Cref{fig:rec_sig_FFT}-left shows the received signal obtained with a modulation frequency of 1~kHz. 
For performing a complete Fourier analysis of the response of the system, the acquisition has been repeated for several modulation frequencies from 100~Hz to 1~MHz. For each acquisition the corresponding fast-Fourier transform (FFT) has been computed (see \cref{fig:rec_sig_FFT}-right for the FFT of the 1-kHz signal). From the FFTs we extracted the modulation response of the system (\cref{fig:freqModBand}), 
the signal-to-noise ratio (S/N) and the harmonic distortion characterizing the received signal (\cref{fig:harm_dist_SNRvsfreq}). 
From the modulation response analysis, a modulation bandwidth of $\sim \SI{300}{\kilo \hertz}$ can be retrieved. This value is perfectly in agreement with the well-known single mode QCLs' frequency modulation bandwidth~\cite{Tombez:2013a}. The S/N has been computed as the ratio between the power of the peak corresponding to the signal and the power level corresponding to the noise floor. On the other hand, the harmonic distortion has been computed as the ratio between the power of the second harmonic and the power of the fundamental frequency peaks of the received signal. These last two parameters have been computed within a bandwidth of 20~kHz, covering a range of values absolutely compatible with a good quality audio transmission. Moreover, a substantial flatness of their values with the modulation frequency clearly emerges, confirming the good performance of the system.

\subsection{Attenuation effects}

For a concrete assessment of the performance of the FSOC system, an evaluation of the impact of optical losses is crucial. Optical losses can be split into two groups: static (or quasi-static) and dynamical. As an example, losses due to optical elements, molecular absorption, aerosol scattering, geometrical losses due to the divergence of the optical beam, belong to the first group. Losses due to atmospheric turbulence belong to the second group. The dynamical losses importantly impact the performance of FSOC systems and are hard to be modeled and simulated. Here we concentrated on the simulation of quasi-static losses effects by placing a variable optical attenuator on the beam path. In particular, we studied how the S/N changes with optical attenuation and we investigated the possibility of recovering it by adapting the electronic gain of the RF amplifier placed at the end of the IBN down-mixing stage (see \cref{fig:QCL-comb_comm_setup}). The chosen modulation frequency is 1~kHz, being representative for audio transmission. 
\begin{figure}[!htbp]
\begin{center} 
\includegraphics[width=0.60\textwidth]{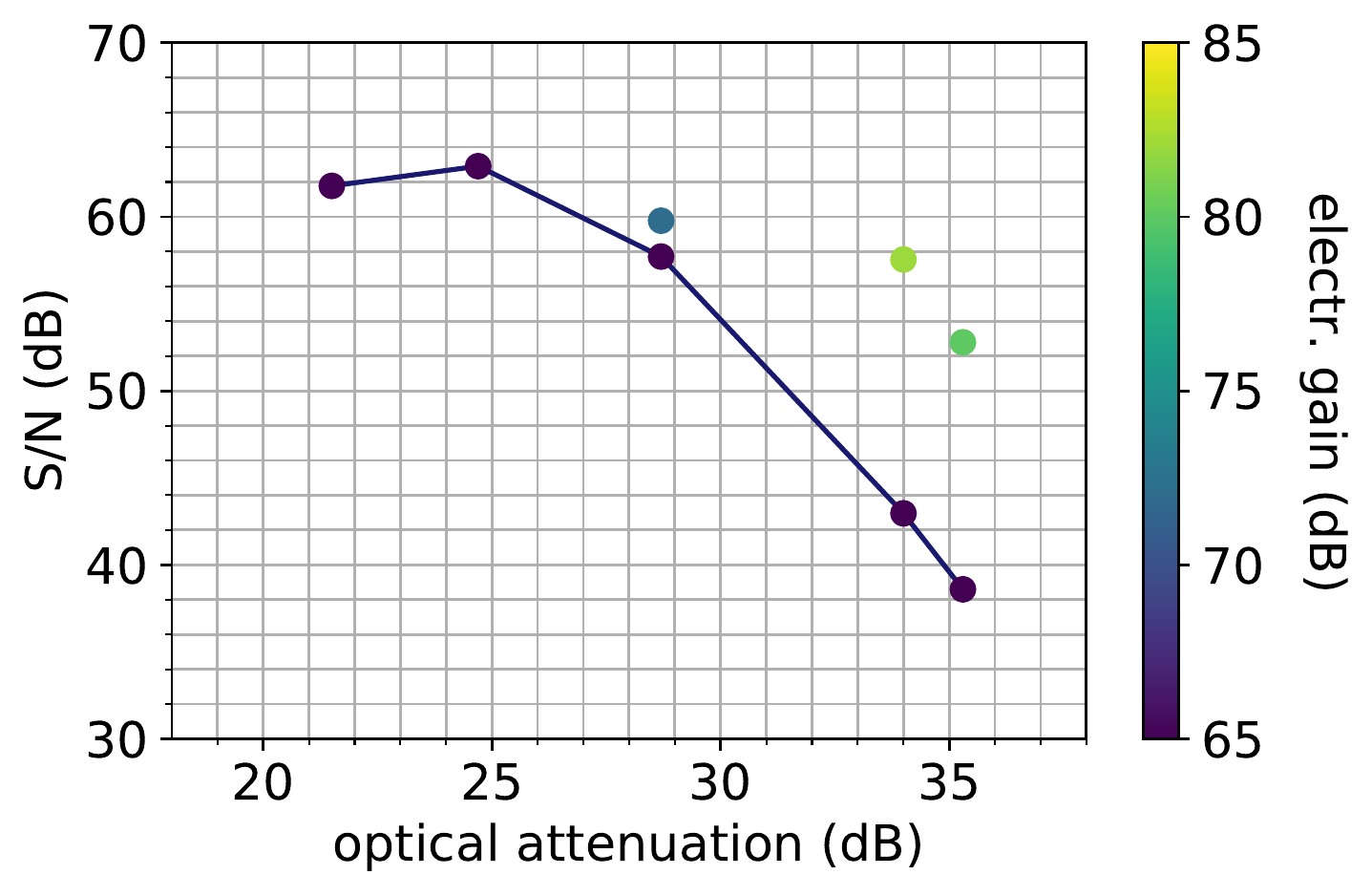}
\caption[]{S/N as a function of optical attenuation for a modulation frequency of \SI{1}{\kilo \hertz}. The variable electronic gain used for recovering the S/N is color coded. } 
\label{fig:SNRvsAttCm}
\end{center}
\end{figure}
In \cref{fig:SNRvsAttCm} the dots lying on the blue curve show a S/N decreasing with increasing optical attenuation (case of fixed electronic gain). On the other hand, the brighter dots show that, when the electronic gain is increased, the S/N can be significantly recovered, up to a value of 14~dB, resulting in a maximum S/N loss of 10~dB over the explored range of optical attenuation (up to 35~dB). 

For giving a practical reference of the scale of optical attenuation values we have investigated, we discuss here the equivalent conditions for an outdoor FSOC. We assume a communication distance of \SI{5}{\kilo\meter} and an optical system optimized for minimizing the geometrical losses, i.e. for collecting all the optical power at the receiver location. We also neglect the possible effects given by atmospheric turbulence. The residual attenuation is generally given by the scattering due to molecules and particles (aerosol). In the spectral region covered by the QCL-comb (around \SI{4.70}{\micro \meter}, see the \nameref{sec:supplementary}), the average molecular absorption value is about \SI{1.22}{\dB / \kilo\meter} \cite{Rothman:2013,HITRANd:2013,HITRANw:2019}, \SI{6.1}{\dB} over 5~km. This means that an additional attenuation of at least \SI{29.0}{\dB} can be tolerated by the system, which corresponds to a visibility range~\cite{note:visibility_def} of \SI{0.88}{\kilo\meter}, given e.g. by light-fog conditions. We remark here that, by using a QCL operating at a more favorable wavelength still within the 4-$\mu$m transparency window, a negligible molecular absorption can be found, allowing an effective communication even with a lower visibility range.

\subsection{Fidelity of the received signal}

\begin{figure}[!htbp]
\begin{center} 
\includegraphics[width=0.48\textwidth]{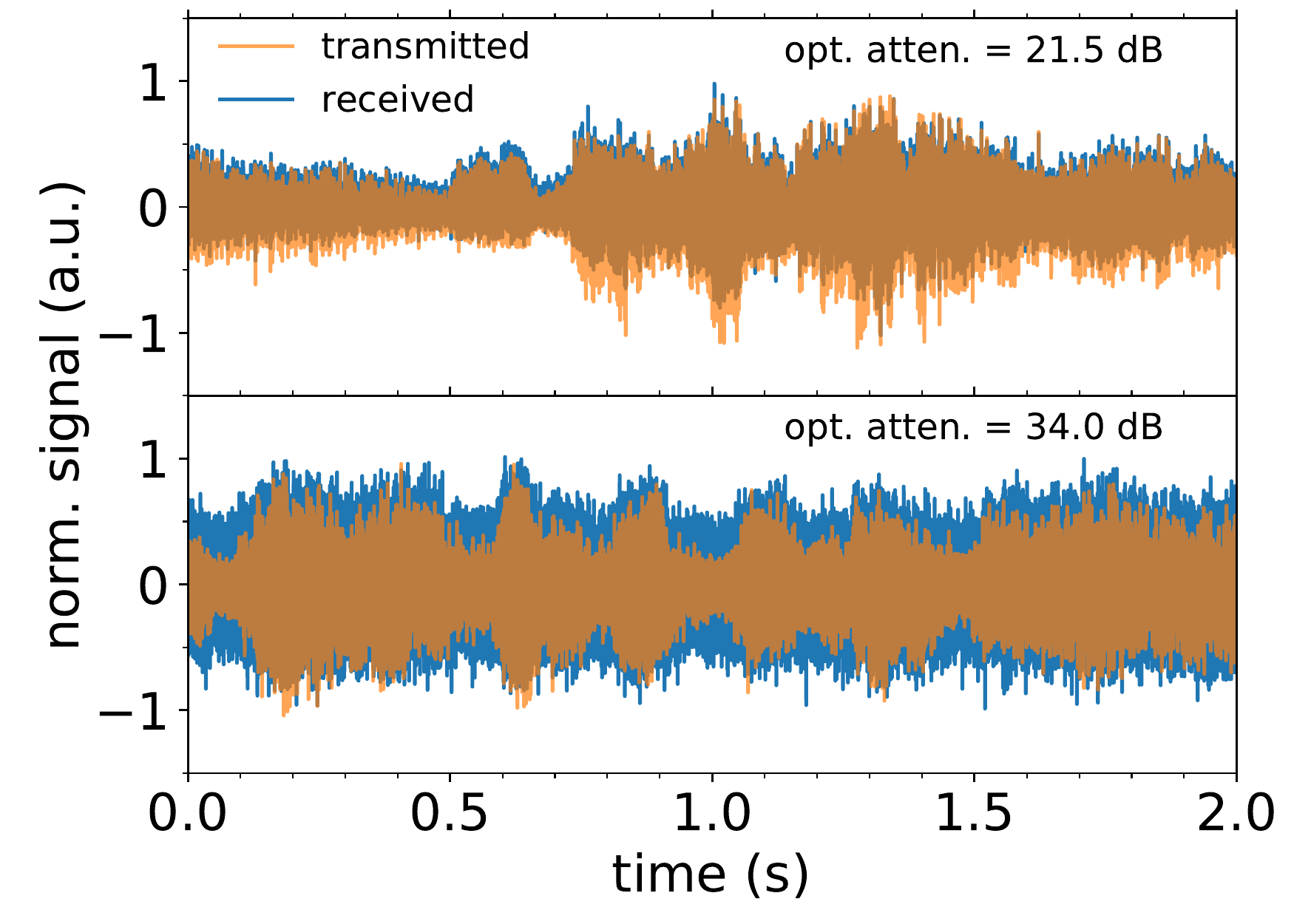} 
\quad
\includegraphics[width=0.48\textwidth]{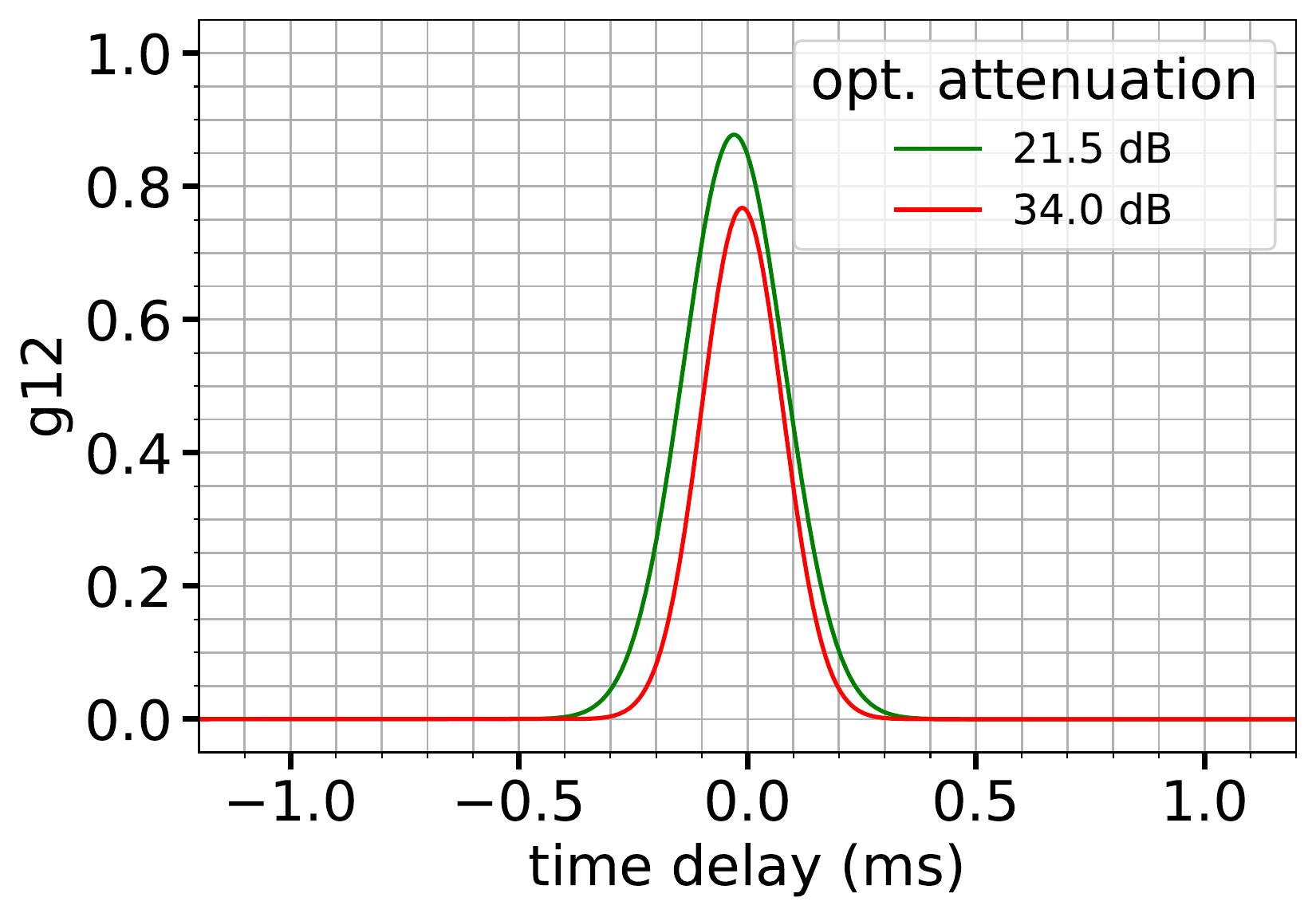}
\caption[]{Left: Sent and received audio signal with two different values of optical attenuation. Right: $g_{12}$ related to the sent and the received signal at the two different levels of optical attenuation. } 
\label{fig:g12_plot}
\end{center}
\end{figure}

The communication system has been tested also for transmission of real audio streams. The MUX has been switched for transmitting audio signals generated by a computer system (CS1 in \cref{fig:QCL-comb_comm_setup}). The sent and the received signal have been acquired simultaneously with an oscilloscope for comparison, with two different levels of optical attenuation (see \cref{fig:g12_plot}-left). 
The received signal looks very similar to the sent one, especially with the lower optical attenuation. For quantitatively estimating the fidelity of the received audio signal, we chose as estimator the first-order correlation between the received and the sent signal ($g_{12}$). The normalized first-order correlation is defined as 
\begin{equation}
g_{12}(\tau) = \frac{\left< S_1(t) \cdot S_2(t-\tau) \right>}{\sqrt{\left< S_1^2 \right> \left< S_2^2 \right>}} \, ,
\label{eq:g12}
\end{equation}
where $S_1$ and $S_2$ are the sent and the received AC signal, respectively, $t$ is time and $\tau$ is the time delay introduced between the two streams. The symbol $\left< ~ \right>$ denotes the averaging process over the arrays and the product $\cdot$ is computed element-by-element. For the computation of the product between the two signals, the second array is shifted of an amount corresponding to $\tau$ and then the two arrays are conveniently cut in order to fit the right number of elements. The arrays are cut as less as possible and the dimension is kept constant over the whole computation. In \cref{fig:g12_plot}-right the $g_{12}$ is represented for two different levels of optical attenuation. The significant value is the maximum of the curve which is 0.87 for the lower attenuation case and 0.75 for the higher attenuation, confirming that, providing an additional electronic amplification, for optical attenuation values within 35~dB the quality of the transmission is not significantly degraded. On the other hand, the width of the curve is related to the time response of the system. The linewidth of $\sim \SI{0.2}{\milli \second}$ corresponds to a low-pass frequency of 5~kHz. This cutoff is probably due to the audio signal generator and is compatible with audio stream reproduction, since voice and music spectra generally fall within this frequency. Some recordings of audio tracks transmitted with the system have been performed for demonstration purposes: See the \nameref{sec:supplementary} for a description.

\subsection{Simultaneous analog FM/digital AM transmission }

\begin{figure}[!htbp]
\includegraphics[width=0.48\textwidth]{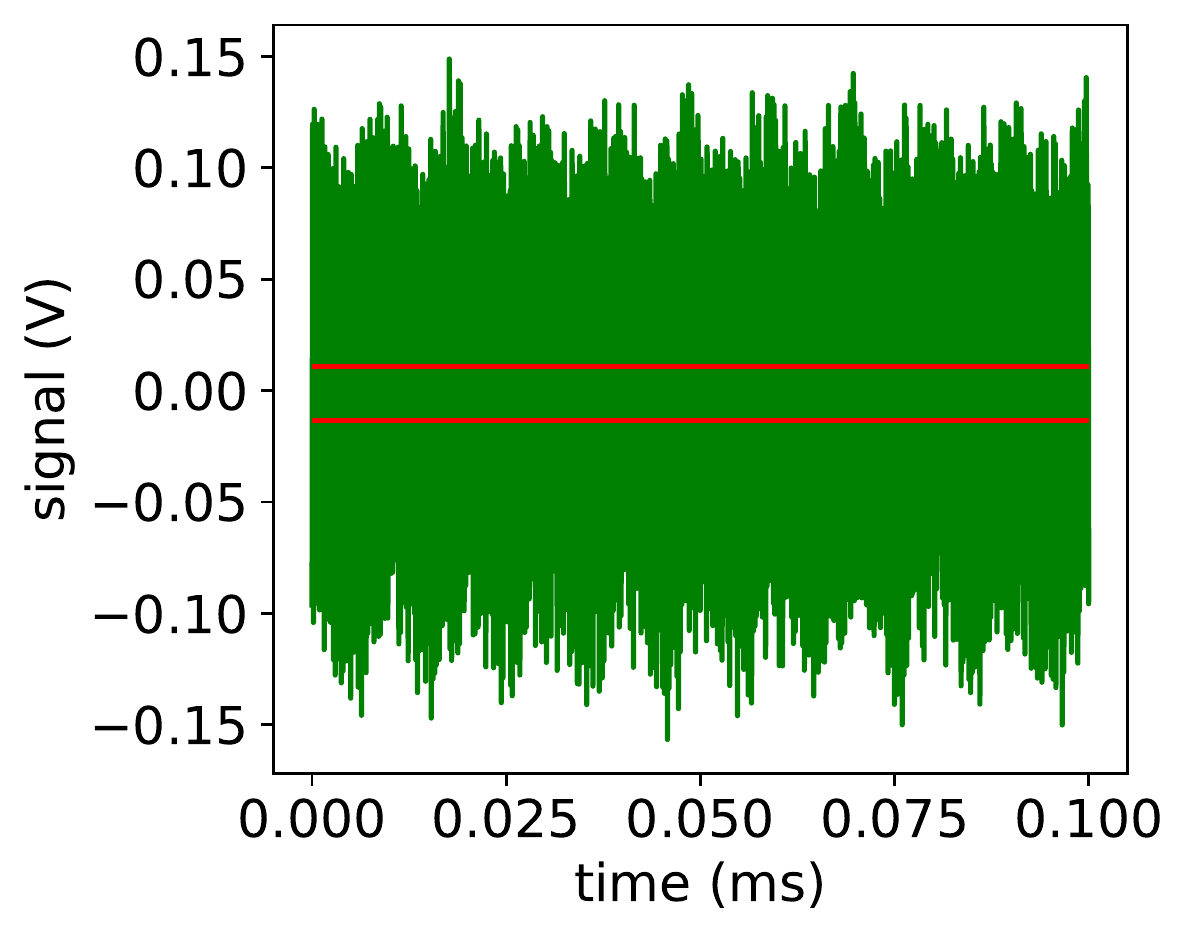} 
\quad
\includegraphics[width=0.48\textwidth]{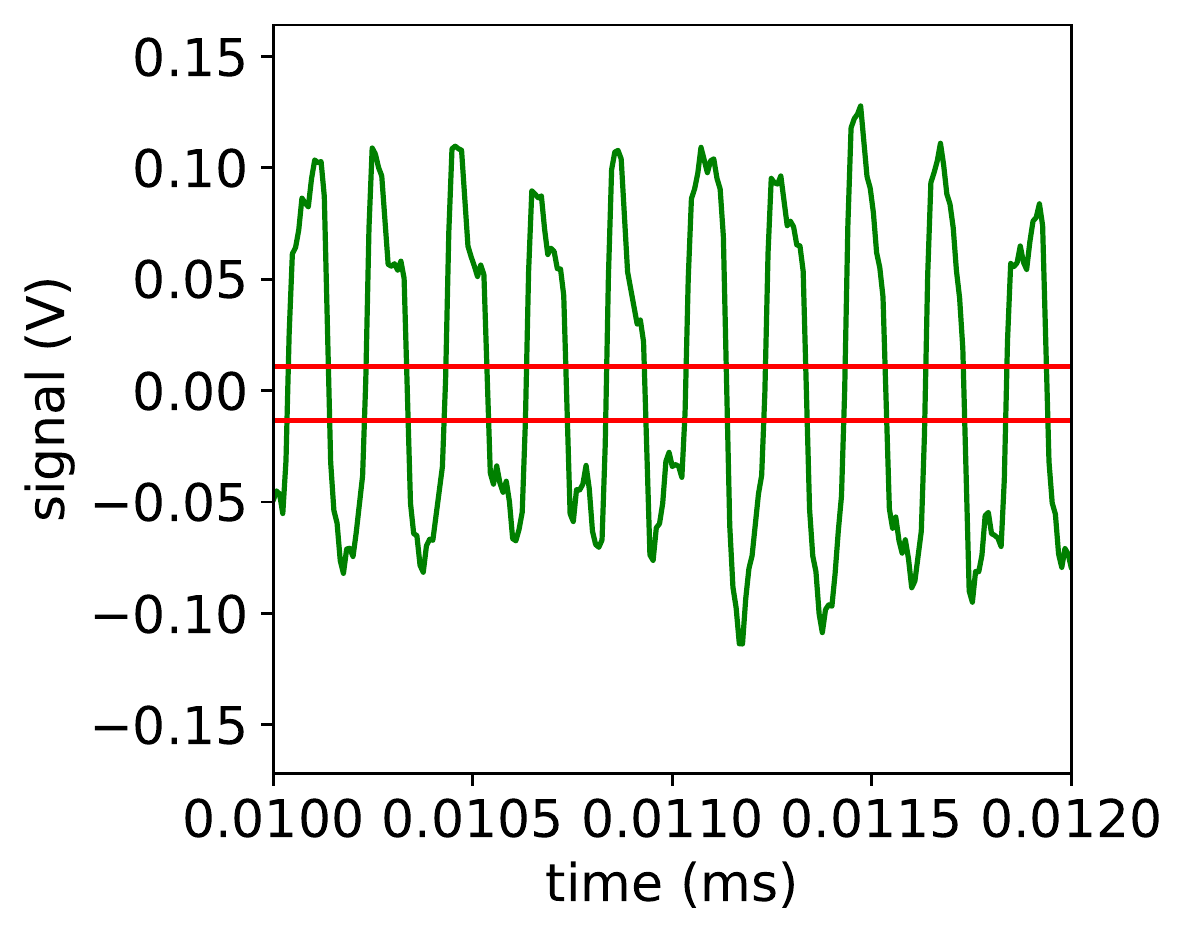}
\includegraphics[width=0.48\textwidth]{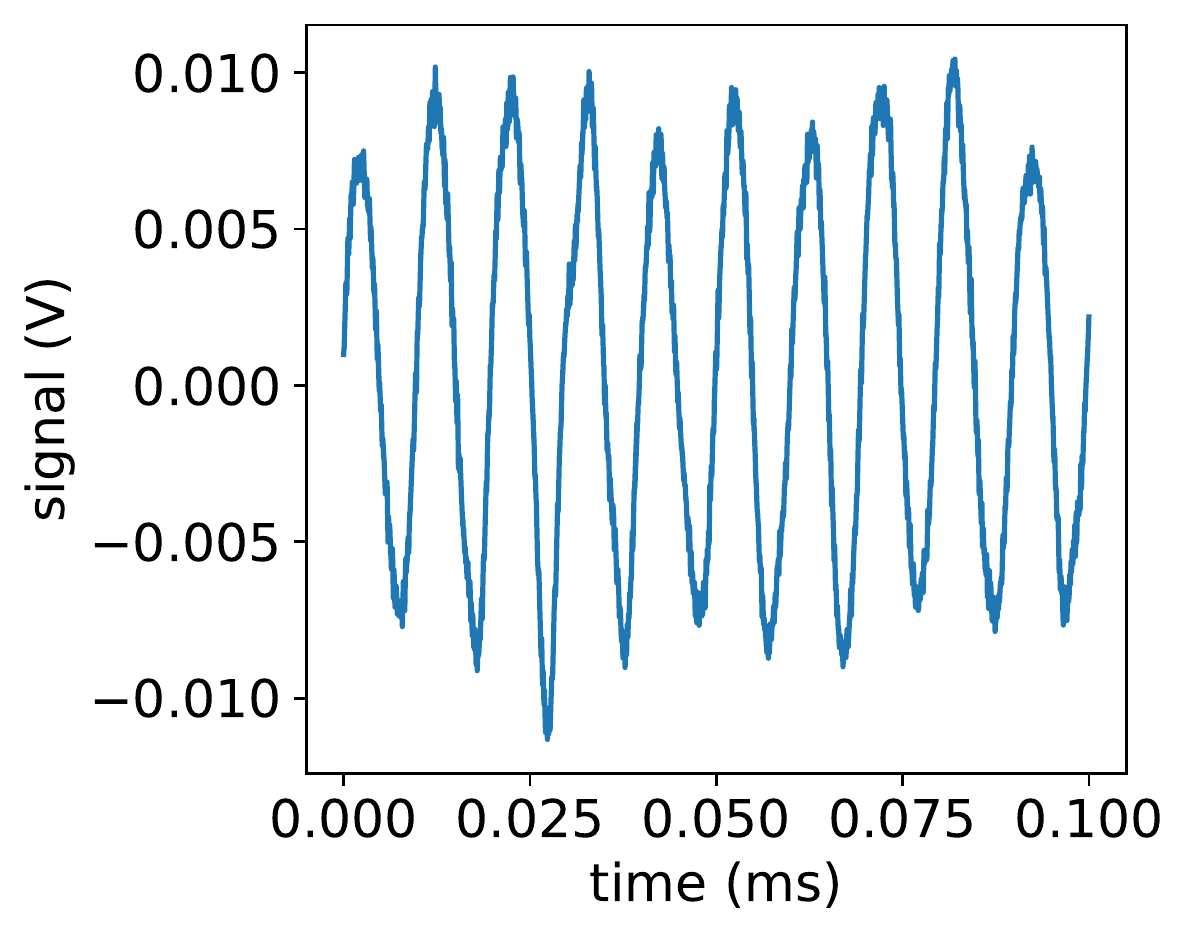}
\quad \quad
\begin{minipage}[b]{0.42\textwidth}
\caption[]{Simultaneous digital AM + analog FM transmission with an optical attenuation of \SI{21.5}{\dB}. Top left: Received square wave. Top right: Square wave zoom. The thresholds set for the digital communication are highlighted. Bottom left: Demodulated sine wave transmitted in parallel with the square wave. }
\label{fig:square_wave}
\vspace{1cm}
\end{minipage}
%
\end{figure}

The realized communication system is also capable of simultaneously transmitting on the same optical channel an independent AM digital signal together with the FM analog signal. The MUX has been set to send in parallel a 100-kHz sine wave (FM of the IBN) and a 5-MHz square wave (AM modulation of the QCL optical power). The results are represented in \cref{fig:square_wave}. 
In this case the optical attenuation is \SI{21.5}{\dB}. The thresholds set for the digital communication are highlighted. The two thresholds have been selected as the maximum and the minimum signal levels where the repetition of the fronts of the signal does differ less than 30\% with respect to its mean value (computed at the zero crossing) over the whole acquisition. The two thresholds define a window of \SI{24}{\milli \volt}, that is more than sufficient for guaranteeing an error-free digital communication. Moreover, from the same figure a good shape of the analog FM signal can be inferred. This configuration has been tested for several values of optical attenuation up to \SI{28.7}{\dB}. The system still maintains a good capability of transmission with a digital window of \SI{4}{\milli \volt} and a good shape of the demodulated analog FM signal. This demonstrates the possibility of exploiting the full FM bandwidth of the system for analog communication (\SI{300}{\kilo \hertz}) together with an independent digital communication at a speed of \SI{5}{\mega S / \second}, essentially limited by the current modulator and not by the QCL internal dynamics.

\section{Conclusion}

In this work, we have demonstrated the possibility of exploiting the intermodal beat-note signal (IBN) spontaneously-generated by QCL-combs for performing point-to-point free-space optical communication in the MIR transparency window around \SI{4}{\micro \meter}. A S/N up to \SI{65}{\dB} has been achieved, with a modulation bandwidth of \SI{300}{\kilo \hertz}. With an optical attenuation up to \SI{35}{\dB} the loss in terms of S/N is less than \SI{10}{\dB}. 
With this system, audio streams of a very good quality have been transmitted, with a fidelity expressed in terms of first-order correlation between the sent and received signal ($g_{12}$) up to 87\%. The transmission has been performed on a single channel (monophonic) for simplicity, but the wide bandwidth characterizing the system is potentially able to host e.g. a 38~kHz stereo subcarrier allowing the transmission of a composite stereophonic signal typically adopted for FM radio broadcasting. 

The system also allows a parallel transmission of an independent digital signal via AM at a speed of \SI{5}{\mega S / \second}. 

The presented proof-of-principle system could be improved in many aspects. First of all, the use of QCLs with optimized growth, fabrication and contacting would strongly increase the AM bandwidth~\cite{Hinkov:2016,Hinkov:2019} and possibly also the FM bandwidth. On the detection and demodulation side, by implementing the detection of the digital signal as modulation of the IBN around \SI{7}{\giga \hertz} in place of the performed direct detection we expect a significant improvement in terms of S/N. In order to fully exploit the advantages of FM communication, a dynamic-range compressor amplifier could be added right before the DLD. This would make the demodulation totally independent of the received signal amplitude turning the contribution of transmission amplitude noise completely negligible. Moreover, more compact and efficient demodulation schemes could be adopted in place of the DLD, e.g. based on a phase-locked loop. 
Finally, alternative transmission schemes able to exploit a balanced detection~\cite{Gabbrielli:2021a} could improve further the S/N of the communication by suppressing common-mode noise.

\section*{Acknowledgements}

The authors gratefully thank Mauro Giuntini and Roberto Concas for useful discussions. 

The authors gratefully thank the collaborators within the consortium of the Qombs Project: Prof. Dr. Jérome Faist (ETH Zurich) for having provided the quantum cascade laser and the company ppqSense for having provided the ultra-low-noise current driver (QubeCL). 

The authors acknowledge financial support by the European Union’s Horizon 2020 Research and Innovation Programme with the Qombs Project [FET Flagship on Quantum Technologies grant n. 820419] ``Quantum simulation and entanglement engineering in quantum cascade laser frequency combs''.



\section*{Disclosures}
The authors declare no conflicts of interest. 

\section*{Data availability}

The data that support the findings of this study are available from the corresponding authors upon reasonable request.

\bibliographystyle{ieeetr_CA}
\bibliography{references.bib}

\clearpage

\renewcommand{\thesection}{S}

\section{Supplementary information}
\label{sec:supplementary}

\subsection{Laser characterization}

\begin{figure}[!htbp]
\begin{center} 
\includegraphics[width=0.48\textwidth]{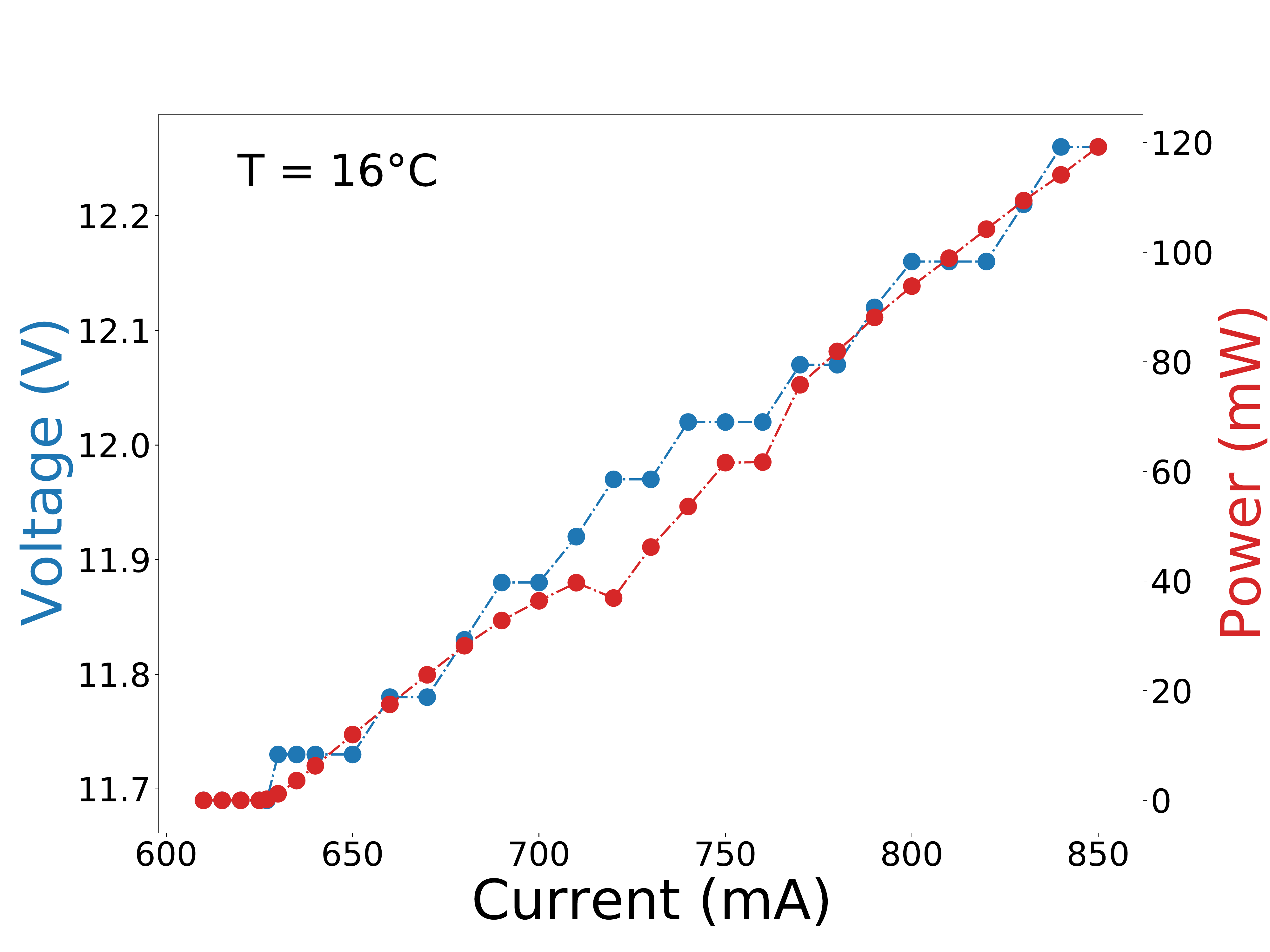}
\quad
\includegraphics[width=0.48\textwidth]{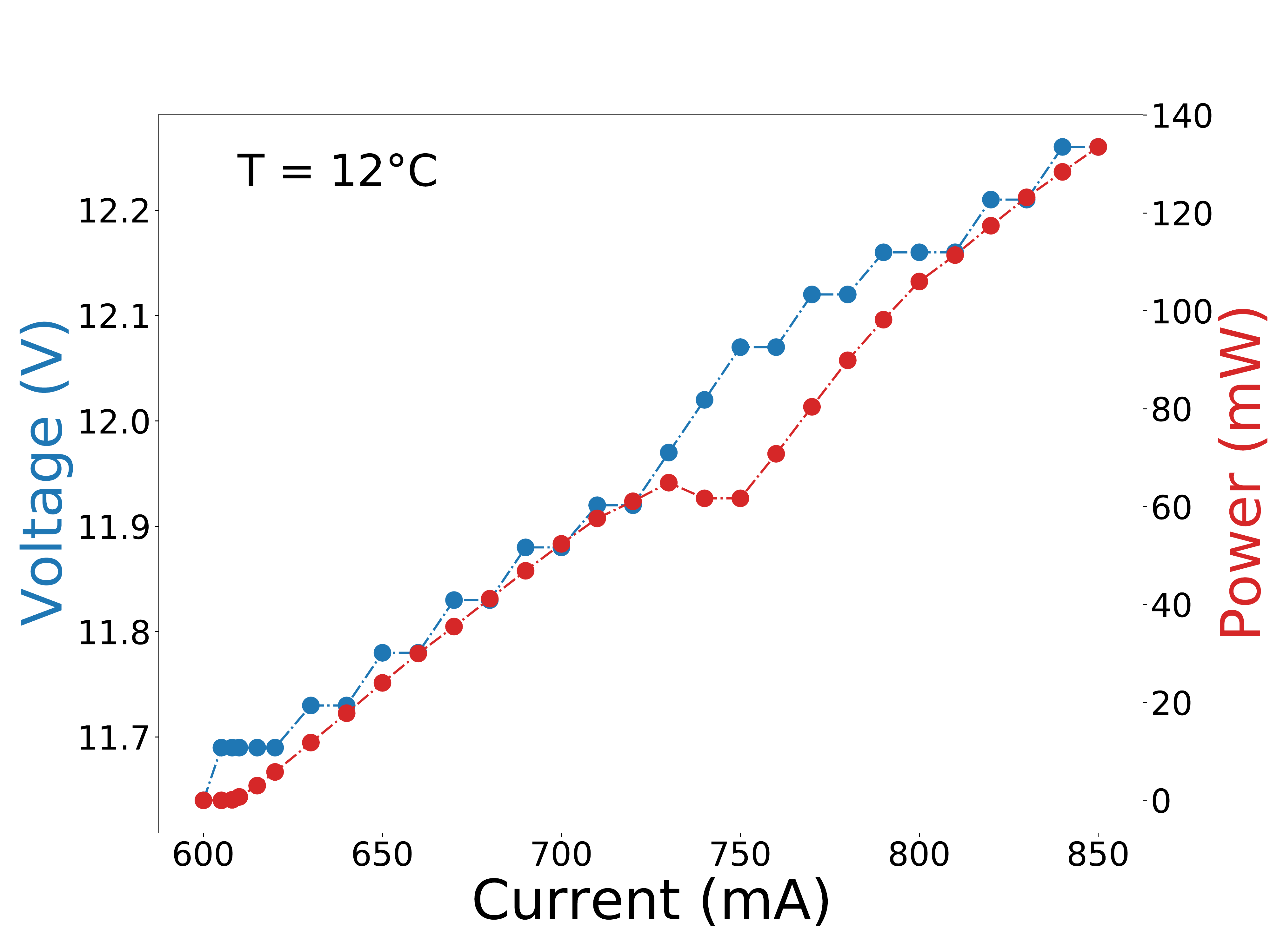}
\caption[]{QCL LIV curves at two different temperatures. } 
\label{fig:LIV}
\end{center}
\end{figure}

The QCL used for this demonstration is characterized by the LIV curves represented in \cref{fig:LIV}. 
Interestingly we note that at $T = \SI{12}{\celsius}$ and $I = \SI{730}{\milli \ampere}$ the power tuning with driving current is minimum, in correspondence of the maximum of frequency tuning of the IBN. This working point is favorable for FM transmission. 

\begin{figure}[!htbp]
\begin{center} 
\includegraphics[width=0.7\textwidth]{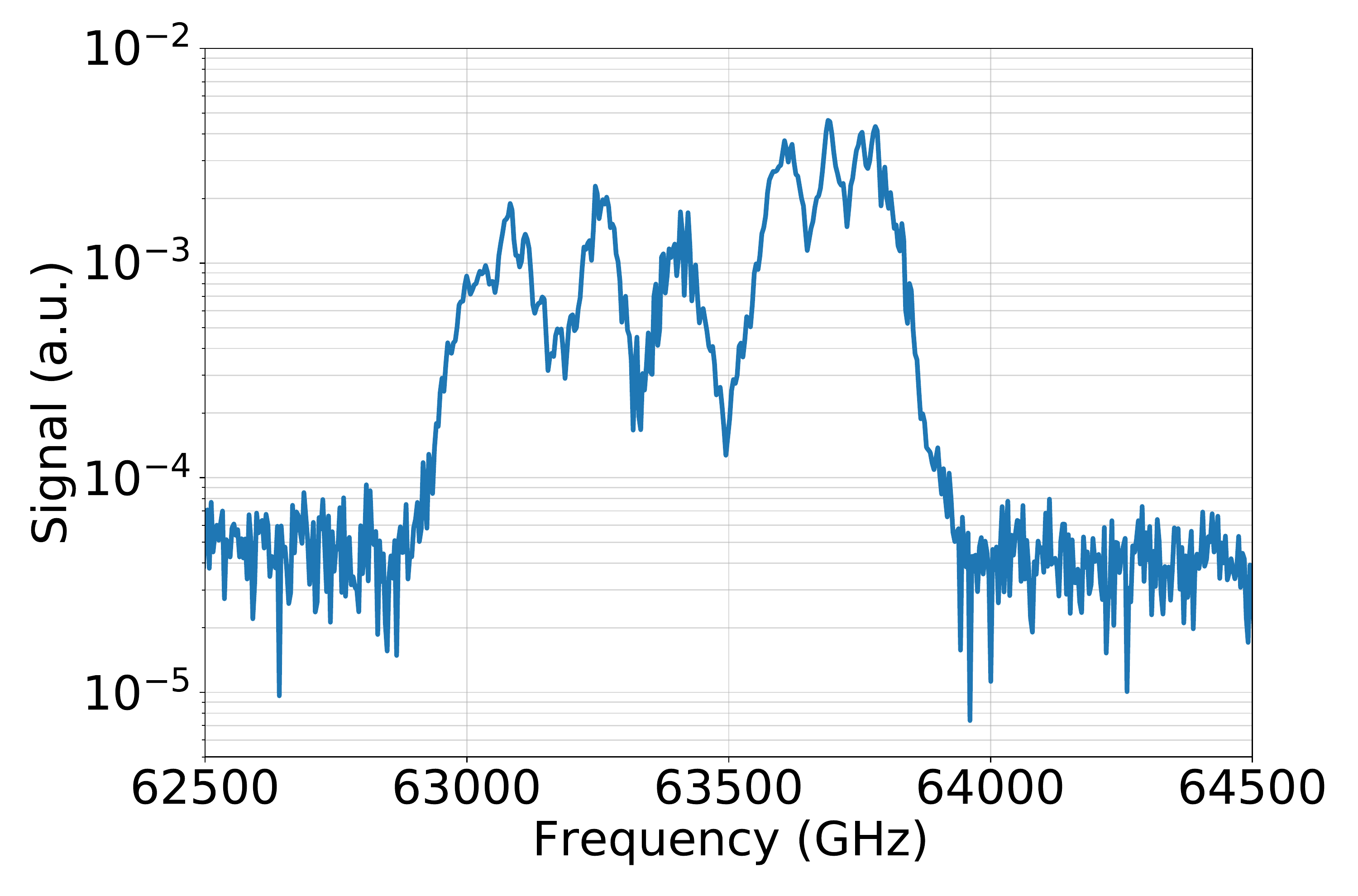}
\caption[]{Spectrum of the QCL-comb operating at $T = \SI{12}{\celsius}$ and $I = \SI{730}{\milli \ampere}$ detected by an optical spectrum analyzer. } 
\label{fig:QCL_spectrum}
\end{center}
\end{figure}

In \cref{fig:QCL_spectrum} the spectrum of the QCL-comb operating at $T = \SI{12}{\celsius}$ and $I = \SI{730}{\milli \ampere}$ is shown. 

\subsection{Delay line frequency discriminator}

The DLD used for the demodulation of the received signal~\cite{Faith:1988} is represented in the blue box in \cref{fig:QCL-comb_comm_setup}. At first, the input signal is divided into two parts. One half is directly applied to an RF mixer, while the second half is delayed by passing through a 22-m-long BNC cable and then applied to the other input of the RF mixer. The role of the RF mixer is to generate a non-zero-average output signal by squaring the interference between the input signal and its delayed version. The output of the RF mixer is low-pass filtered by LPF1 for removing the carrier. In this way, at the output of the DLD a DC signal proportional to the input frequency is generated. To obtain the optimal working point of the DLD and minimize the amplitude noise while maximizing the signal frequency-to-amplitude conversion, a thorough characterization of the DLD itself is required. For the characterization a sinusoidal signal is applied to the DLD input: 
\begin{equation}
V_{in}=V_0\cdot\sin(\omega ~ t) \, ,
\label{eq:dld1}
\end{equation}
with $\omega=2\pi f$, $V_0=\SI{0.5}{\volt}$ and $f$ is the frequency spanning in the range between 46~MHz to 64~MHz. 
\begin{figure}[!htbp]
\begin{center} 
\includegraphics[width=0.6\textwidth]{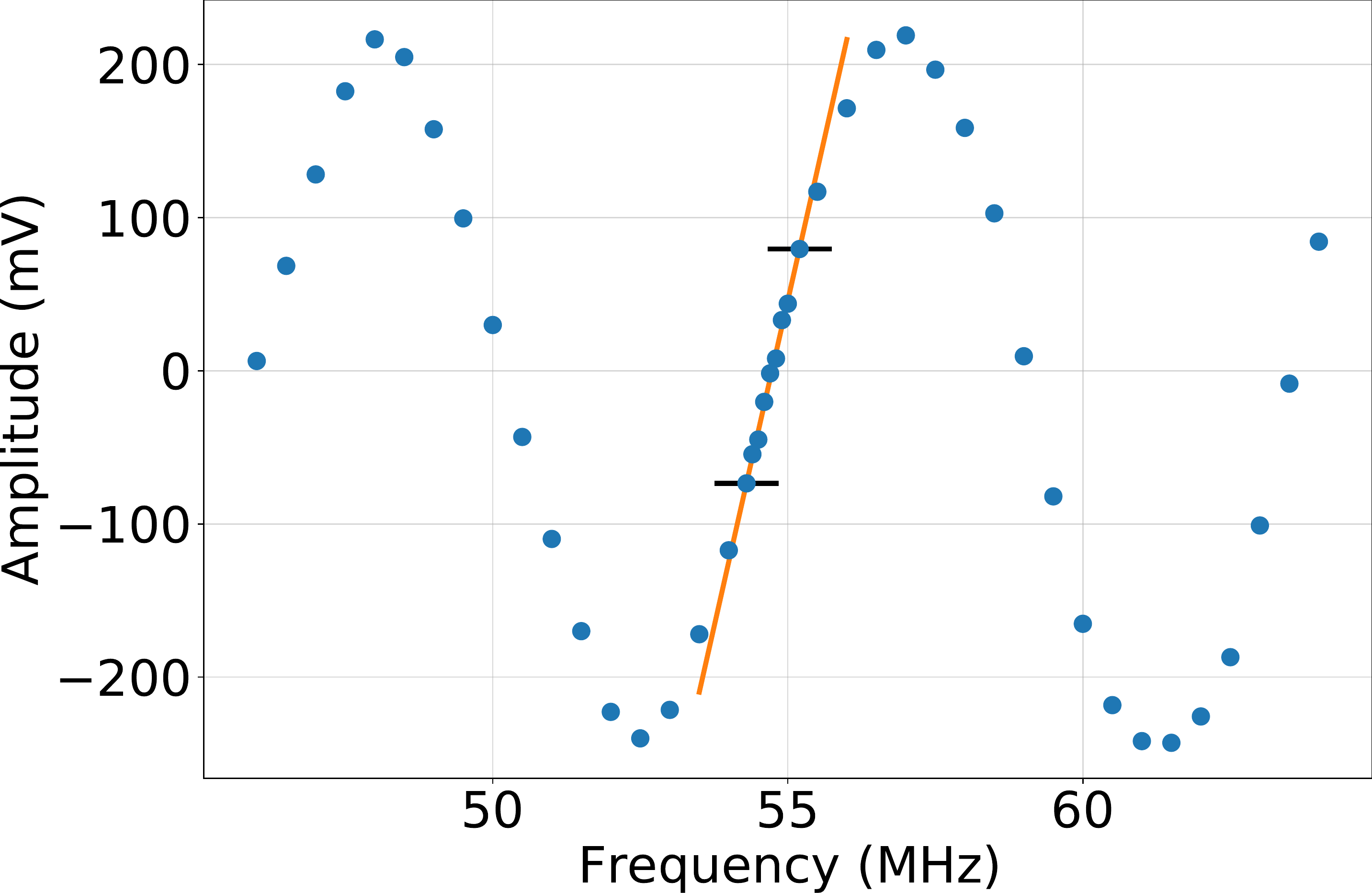}
\caption[]{DLD frequency response. } 
\label{fig:DLDcharacterization}
\end{center}
\end{figure}
In \cref{fig:DLDcharacterization} the transfer function of the DLD is represented. As the frequency changed, the DC output value was measured. The period $i_{d}$ of the DLD response function is proportional to the delay of the line $(\tau)$: 
\begin{equation}
\tau = \frac{1}{i_d } \, ,
\label{eq:dld2}
\end{equation}
with $i_d = \SI{9}{\mega \hertz}$. The speed of propagation in the line is:
\begin{equation}
v \simeq \frac{2}{3}\cdot 3 \cdot 10^8 \SI{}{\meter\per\second} \, .
\label{eq:dld4}
\end{equation}
The length of the delay line can be therefore computed:
\begin{equation}
L = \frac{v}{i_d } \simeq \SI{22.2}{\meter} \, .
\label{eq:dld3}
\end{equation}
The obtained value is in good agreement with the measured one. The optimal working point of the DLD is for an input frequency of \SI{54.6}{\mega \hertz}, where the frequency response has a linear zone with the maximum slope. The maximum slope allows for an optimized frequency-to-amplitude conversion and therefore an optimized demodulation capability. For this reason the LO in the previous down-mixing stage (see the green box in \cref{fig:QCL-comb_comm_setup}) has been set to have a down-mixing of the IBN from 7.06~GHz right to 54.6~MHz.

\subsection{Audio tracks recordings transmitted with the system} 

For demonstrating the system performance with real signals, three audio tracks have been transmitted and recorded. The system supports a single channel, therefore the recordings are monophonic. In particular, the CS2 audio board has been used for acquiring the signal employing the open-source software Audacity. No effects have been used to enhance the audio quality. The recorded audio tracks, listed below, can be found in the Ancillary files page related to this article. 
\begin{enumerate}
    \item \emph{The trumpet shall sound} from Messiah by George F. Handel (extract) -- with optical attenuation of 21.5~dB and 34.0~dB 
    \item \emph{Der H\"olle Rache} from The Magic Flute by Wolfgang A. Mozart (extract) -- optical attenuation 21.5~dB 
    \item \emph{Spunta la luna dal monte} by Pierangelo Bertoli and the Tazenda (extract) -- optical attenuation 21.5~dB 
\end{enumerate}

\end{document}